\documentclass{llncs}

\input{psfig.sty}
\usepackage{latexsym}
\newlength{\symbwidth}

\newcommand{\ndres}{\mbox{$\lhd$\settowidth{\symbwidth}{$\lhd$}\hspace*{-\symbwidth}$-$}}
\newcommand{\nrres}{\mbox{$\rhd$\settowidth{\symbwidth}{$\rhd$}\hspace*{-\symbwidth}$-$}}

\font\msx=msam10
\font\msy=msbm10

\newfam\msxfam \textfont\msxfam=\msx
\newfam\msyfam \textfont\msyfam=\msy

\def\famletter#1{\ifcase #1 0\or 1\or 2\or 3\or 4\or 5\or 6\or 7\or
        8\or 9\or A\or B\or C\or D\or E\or F\fi}

\edef\fx{\famletter\msxfam}

\def\p#1{\mathrel{\ooalign{\hfil$\mapstochar\mkern 5mu$\hfil\cr$#1$}}}

\let\mc=\mathchardef

\let    \fun            \rightarrow
\def    \pfun           {\p\fun}

\mc     \inj            "3\fx1A

\def    \psurj          {\p\surj}
\mc     \surj           "3\fx10

\def    \bij            {\surj\hspace*{-2.58ex}\inj}

\begin{document}

\pagestyle{empty}

\title{Reuse of Specification Patterns \\
with the B Method}

\author{Sandrine Blazy \and Fr\'{e}d\'{e}ric Gervais \and R\'{e}gine Laleau}

\institute{Institut d'Informatique d'Entreprise, Laboratoire CEDRIC, \\
18, All\'{e}e Jean Rostand, F-91025 \'{E}VRY Cedex, France \\
\email{\{blazy, gervais, laleau\}@iie.cnam.fr}}

\maketitle

\begin{abstract}
This paper describes an approach for reusing specification patterns. 
Specification patterns
are design patterns that are expressed in a formal specification language.
Reusing a specification pattern means instantiating it or composing it with 
other specification patterns.
Three levels of composition are defined: juxtaposition, composition with 
inter-patterns
links and unification.
This paper shows through examples how to define specification patterns in B, 
how to reuse them
directly in B, and also how to reuse the proofs associated with
specification patterns.

\end{abstract}

\section{Introduction}

Component-based development is a technology that has been widely used for 
many years during the implementation phase of the software life cycle. 
This technology has also been adapted to the design phase. Design patterns 
expressed in UML are frequently reused in order to simplify the design 
of new applications. The most famous design patterns are called the Gang of 
Four (or GoF) patterns \cite{gamm95}.
During the development of a new application, most designers refer to the GoF 
patterns, even if there is no precise definition of these patterns. 
This lack of a formal definition allows designers to adapt freely the GoF 
patterns to their needs and has contributed to the success of the GoF patterns.

As formal specifications are now well-known in industry, the reuse of formal 
specifications based on design patterns becomes a challenging issue. 
Reusing a formal specification means firstly to formally define components 
of formal specifications (or specification patterns). Secondly, it means 
to define how to combine components together in order to build a new 
application. Other problems such as the definition of a library for 
storing the components must also be solved.
A few works address the problem of defining specification patterns. Each 
of these works define also a specific way to combine specification 
patterns together but there is no consensus on the definition of a 
specification pattern or on the combination of patterns.

We use the B language to formally specify the notion of specification 
pattern, and several ways to combine specification patterns together.
Our approach aims at helping the designer to firstly formally specify a 
new application that reuses design patterns and 
secondly to assist him with a tool. We have chosen the B language for 
the following reasons: 
\begin{itemize}
\item Where B is already being used, then there is no need to learn a new 
formalism to define and reuse specification patterns.
\item B is supported by tools that validate the specification. We will 
use them to validate the definition of specification patterns and
the different reuse mechanisms.
A designer will thus reuse not only pieces of formal specifications but 
also proofs concerning 
these pieces of formal specifications. 
\end{itemize}

This paper describes our approach through examples and is organised as follows.
The next section is an introduction to patterns. Section 3 deals with 
a state of the art about the reuse of specification patterns with formal 
methods. Section 4 is a discussion of the notion of reuse in B: our 
approach is presented and illustrated by an example. Finally, we conclude 
this work in Sect. 5 with the perspectives and the limits of this 
approach.

\section{About Patterns}

The aim of this section is to present in an informal way how to specify an 
application with UML patterns. This section identifies also several ways 
to reuse the patterns. 

\subsection{Examples of Patterns}

Two examples of design patterns are presented in this section. They will be 
used in the remainder of the paper. 

Figure 1 presents with the UML notation the class diagram of the 
{\bfseries Composite} design pattern \cite{gamm95}. This pattern is a 
solution to describe the tree structure of a composite, which is an object 
composed of several other objects. Two classes are defined to represent 
composite objects (\texttt{Composite} class) and basic objects (\texttt{Leaf} 
class). An abstract class called \texttt{Component} represents both 
composite and basic objects. An association is defined between 
\texttt{Composite} and \texttt{Component} classes with the \texttt{father} 
and \texttt{children} roles. A \texttt{Composite} can have 
several \texttt{Component} children which can be \texttt{Leaf} or 
\texttt{Composite} objects.

\begin{figure}
\centerline{\psfig{figure=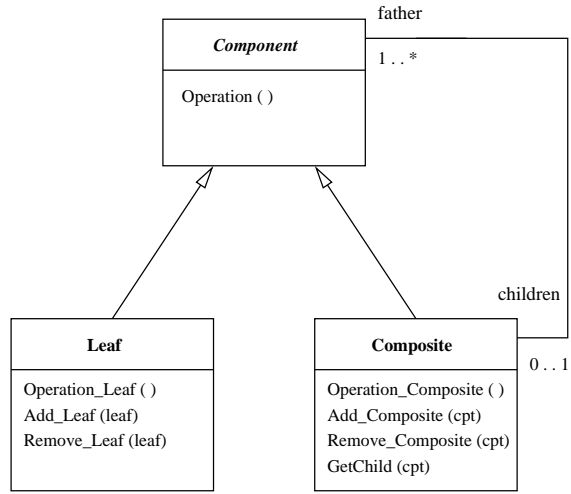,height=6.5cm}}
\caption{Class diagram of the {\bfseries Composite} design pattern}
\label{fig:composite}
\end{figure}

Operations (methods) are defined in the different classes. The set of 
components of a composite object is given by the \texttt{GetChild} method. 
\texttt{Operation} is a generic operation and deals with both leaf and 
composite objects. This operation is redefined in the \texttt{Composite} 
and \texttt{Leaf} classes by \texttt{Operation\_Composite} and 
\texttt{Operation\_Leaf}.

Figure 2 presents the class diagram of the {\bfseries Resource Allocation} 
pattern described in \cite{fowler}. Four classes are defined. The 
\texttt{Resource} class represents a resource to allocate. A resource 
provides facilities, represented in this pattern by the 
\texttt{ResourceFacility} class. A resource is allocated to job 
occurrences represented by the \texttt{JobOccurrence} class. Finally, 
\texttt{JobCategory}, which stands for the job categories, is linked to 
the last two classes. The requirements of the resource facilities are 
supported only by specific job categories. A job is represented by an 
association between \texttt{JobOccurrence} and \texttt{JobCategory}. 

\begin{figure}
\centerline{\psfig{figure=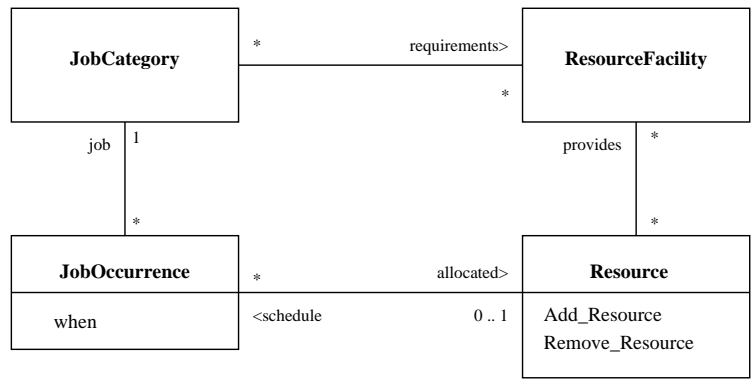,height=5.0cm}}
\caption{Class diagram of the {\bfseries Resource Allocation} pattern}
\label{fig:resalloc}
\end{figure}

\subsection{Design by Reuse}

To motivate the need for reuse of specification patterns, we use the simple 
example of designing in UML the allocation of directories to secretaries. 
A directory is composed of files and other directories. Figure 3 gives a 
solution obtained by ``instantiating'' both patterns {\bfseries Composite} 
and {\bfseries Resource Allocation}. In the first pattern, a directory is 
considered as a \texttt{Composite} object and a file as a \texttt{Leaf} 
object. The \texttt{Component} class is renamed as \texttt{Element} (note 
that we could have kept the name \texttt{Component}). In the pattern 
{\bfseries Resource Allocation}, an element is considered as a 
\texttt{Resource} object and a secretary as a \texttt{JobCategory} object.

\begin{figure}
\centerline{\psfig{figure=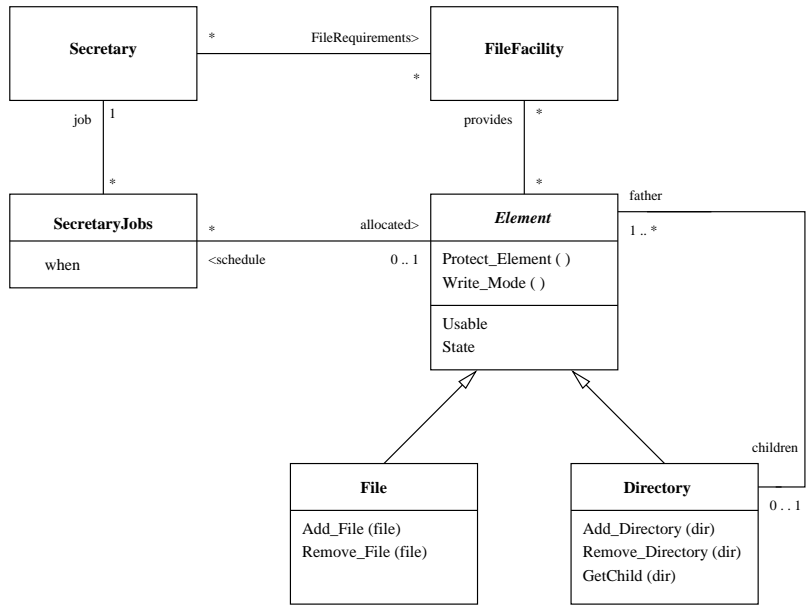,height=8.0cm}}
\caption{An example of design by pattern reuse}
\label{fig:reusexample}
\end{figure}

Compared to both original patterns, some operations are renamed. For 
instance, \texttt{Add\_Composite} becomes \texttt{Add\_Directory} in Fig. 3. 
Other operations such as \texttt{Protect\_Element} and \texttt{Write\_Mode} 
are created once the patterns have been instantiated. New variables
called \texttt{State} and \texttt{Usable} are introduced to describe the
protection mode of a file or a directory which is declared as
usable.

\subsection{Main Operations on Patterns}

Two kinds of operations can be distinguished: pattern definition and 
mechanisms for reusing patterns.

\paragraph{Pattern Definition.}

Two ways exist to define a new pattern. The pattern can either be defined 
``ex nihilo'' or it can be deduced from existing patterns: this kind of 
reuse is aiming to create new patterns. In this case, mechanisms are 
defined at the pattern level in order to link and compare several patterns.  

\paragraph{Reuse Mechanisms.}

Three basic mechanisms exist for reusing patterns in order to design an 
application: instantiation, composition and extension \cite{prieto}. They 
allow patterns to be adapted to the design in progress. 
\begin{itemize}
\item The \textit{instantiation}, also called imitation, is a mechanism 
which allows different elements of a pattern to be renamed. 
\item The \textit{composition} mechanism, also called integration mechanism, 
associates two or more patterns. Different kinds of association may be defined.
\item The last reuse mechanism, called \textit{extension}, allows new 
elements to be added to existing patterns or existing elements to be 
removed or modified.
\end{itemize}

Formal specification languages are very well adapted to define all these 
mechanisms. In the following sections, we present different existing 
propositions, before describing our solution.

\section{State of the Art of Pattern Reuse with Formal Methods}

Up to now, our research works deal with the use of formal methods in order 
to verify functional properties of systems (especially information
systems \cite{ASE00}). Taking into account dynamic properties is a work 
in progress \cite{frappier}. Therefore, in this paper, we do not consider 
event or temporal logic based formal methods such as \cite{lano,mikkonen}.

We have chosen to present four examples of formal methods that 
have been used to formalise design patterns
and their reuse. For each of them, the following criteria
are studied: presentation of the approach, pattern definition,
 available reuse mechanisms and their definition, existence
of tools, usability of the approach, in particular the
required mathematical background and the level of abstraction patterns 
that are defined and reused. More details can be found in \cite{gervais}.

\subsection{UML-B Patterns Reuse}

In \cite{marcano,levy}, design patterns are specified in both UML and B.

\paragraph{Pattern Definition.}

Patterns are defined in two ways. In \cite{marcano}, patterns are
defined ``ex nihilo''. The UML diagram of a GoF pattern is
translated from UML to B using a set of rules. Thus the B specification
can be proved, using a B tool. In \cite{levy}, a pattern can be defined as 
a refinement (also called specialisation) of a more generic pattern, 
thanks to the refinement mechanism of the B method.

\paragraph{Reuse Mechanisms.}

The instantiation of a pattern is defined on the UML description of a 
pattern. The
resulting diagram is again translated to B with the set of
rules. This process has not been
formalised, so the resulting B specification must be proved again.

\paragraph{Tools.}

A tool based on the B and UML environments has been
developed. No specific tools are defined to implement the reuse.

\paragraph{Usability.}

This approach requires knowledge of UML and B. Pattern reuse is
defined on the UML representation. Consequently, reuse mechanisms
cannot be formally defined. Proofs associated with a pattern cannot be 
directly reused
to achieve the proofs of its instantiation. However the definition
of patterns by refinement of other patterns is quite interesting and
original.

\subsection{Frameworks in Catalysis}

Catalysis \cite{LauOrnaghi99} is a component-based specification method. 
Framework is
the name used for pattern in Catalysis. The idea is to
formally specify the frameworks by adapting the existing Catalysis
features.

\paragraph{Pattern Definition.}

A specification framework is defined with algebraic specifications
using many-sorted first order logic. Axioms concerning the
framework are specified in Catalysis with first order
logic predicates. Rather than considering the initial model
traditionally used in algebraic specifications as the referring
model for the framework semantics, the theory is here represented
by an isoinitial model which is very interesting, especially
for proving formulas using negation, because it preserves the
negation property contrary to the initial model\footnote{In an initial
theory, the falsity of a ground atom $A$ corresponds to the non-provability
of $A$, while in an isoinitial theory, it corresponds to the provability of
$\neg A$.}. Nevertheless, the existence of the isoinitial model cannot 
be always guaranteed and axioms must be added to ensure its existence.

\paragraph{Reuse Mechanisms.}

The instantiation is implemented by parameterisation. A
parameter is a sort or a relation between sorts. Additional axioms
involving parameters are defined. The axioms must be satisfied by
the instantiation of parameters.

The frameworks composition is likewise. A parameterised framework
$F_{1}$ may be composed with a parameterised framework $F_{2}$ if the involved
axioms are satisfied. Firstly, a renaming map is defined from
$F_{1}$ to $F_{2}$. The aim is to link each element of the $F_{2}$
signature to one element of the $F_{1}$ signature. Then the
framework resulting from the composition of $F_{1}$ and $F_{2}$ is
characterised as follows~: its signature is the renamed signature of
$F_{1}$ and its axioms are the union of the axioms of $F_{1}$ and
$F_{2}$. Note that this composition is associative but not commutative.

\paragraph{Tools.}

No other tools than those existing in the Catalysis approach are used.

\paragraph{Usability.}

This approach formally defines the notions of framework
specification, instantiation and composition. The choice of
adapting an existing approach (Catalysis) is interesting because
it avoids the introduction of new languages or tools. The isoinitial theory
chosen for the framework semantics
simplifies the soundness proof. However, additional axioms must be
defined in order to ensure that an isoinitial model exists.
Furthermore, it requires knowledge about algebraic specification
and about isoinitial theories.

\subsection{Pattern Specification with RSL}

The idea of this approach is to formalise with RSL (RAISE
Specification Language) \cite{flores00oodgof} design patterns 
described by UML diagrams. RSL is based on VDM and an algebraic
specification language.

\paragraph{Pattern Definition.}

A formal model of pattern is defined. It represents a pattern as a
head, a structure and a list of collaborations. The head section
describes the name, purpose and scope of the pattern. The pattern
structure specifies the classes and associations in RSL. In the
last section, the list of collaborations specifies constraints on
the order of operation calls.

\paragraph{Reuse Mechanisms.}

The instantiation mechanism is implemented by a renaming map.
However, a pattern can be instantiated only once. The resulting 
RSL specification can then be extended by new elements.
The composition mechanism is only illustrated through examples.

\paragraph{Tools.}

To our knowledge, no tool has been implemented.

\paragraph{Usability.}

Patterns for oriented-object programming have been studied
extensively in this approach. That is why a lot of oriented-object
programming features have been specified, thus giving
specifications described at a low level of abstraction.

\subsection{LePUS, a Language for Patterns Specification}

The idea in this approach is the definition of a new formal
language, LePUS (LanguagE for Patterns Uniform Specification) 
\cite{eden-lepus},
dedicated to the specification of patterns. This language is based
on higher order monadic logic (HOML).

\paragraph{Pattern Definition.}

The aim is to formalise the GoF patterns. The main properties of
these patterns are expressed by higher order predicates built on
higher order sets of classes and methods. A pattern is then
represented by a conjunction of predicates.
New patterns can be defined from existing ones by projection
of the higher order sets. Roughly speaking, a projection consists in
reducing the order of a set. Thus, these patterns are more
concrete than the original ones.

\paragraph{Reuse Mechanisms.}

The instantiation of a pattern consists in instantiating the sets involved 
in the pattern predicate.

\paragraph{Tools.}

To our knowledge, no tool has been implemented.

\paragraph{Usability.}

This approach needs a strong background in mathematics.
Moreover, this very formal method is difficult to apply, because
the user must exhibit classes, methods and higher order sets
in order to define an instance of a pattern. Defining a higher order 
set means defining
all the sets with a lower order level and so on up to classes and methods.

\subsection{State of the Art: Conclusion}

Table 1 is a comparison between the four above-mentioned approaches 
and our approach, presented in the next section.

\begin{table}
\caption{Formal approaches for pattern reuse}
\begin{center}
\begin{tabular}{llllll}
\hline \noalign{\smallskip}
Approach & UML-B & Framework & RSL & LePUS & Ours \\
\noalign{\smallskip}
\hline
\noalign{\smallskip}
Section & 3.1 & 3.2 & 3.3 & 3.4 & 4 \\
\noalign{\smallskip}
\hline
\noalign{\smallskip}
Specification & UML and & algebraic & VDM & HOML & B\\
models: & B machines & specification & & & \\
\noalign{\smallskip}
\hline
\noalign{\smallskip}
Direct & UML to B & isoinitial & RSL & conjunction & B machine \\
Definition & translation & theory & formal& of predicates & \\
of Pattern: & & & model & & \\
\noalign{\smallskip}
\hline
\noalign{\smallskip}
Pattern & refinement & no & no & projection & no \\
Definition & & & & & \\
by Reuse: & & & & & \\
\noalign{\smallskip}
\hline
\noalign{\smallskip}
Instantiation: & UML & parameters & map & parameters & inclusion \\
 & & & & & and renaming \\
\noalign{\smallskip}
\hline
\noalign{\smallskip}
Composition: & UML & not & no & no & inclusion and \\
 & & commutative & & & new invariant \\
\noalign{\smallskip}
\hline
\noalign{\smallskip}
Extension: & UML & no & yes & no & inclusion and \\
 & & & & & refinement \\
\noalign{\smallskip}
\hline
\noalign{\smallskip}
Tools: & UML and & Catalysis & no & no & B tools \\
 & B tools & & & \\
\noalign{\smallskip}
\hline
\noalign{\smallskip}
Usability & $+$ & $+$ & $-$ & $--$ & $+$ \\
\noalign{\smallskip}
\hline
\end{tabular}
\end{center}
\end{table}

The LePUS approach is the most advanced specification
method for defining and instantiating design patterns, but it is also
the most difficult to apply. The instantiation notion is
defined in all the approaches: by the UML instantiation mechanism 
for UML-B, by
parameterised frameworks for Catalysis, by a renaming map for RSL
and by an instantiation of parameters for LePUS. The composition
mechanism seems to be more difficult to define, since only a non
commutative operation is defined for frameworks. The extension is
only treated by RSL and UML-B. Even if the UML-B approach seems to be 
close to our
work, it differs significantly since the reuse mechanisms
are not formally defined.

\section{An Approach for Reusing Patterns in B}

The aim of this section is to investigate the ability of B to
specify patterns and the different reuse mechanisms. We have chosen to 
consider the B
language as it is. This means that we want to define the  reuse mechanisms only
with the different B mechanisms such as refinement, inclusion, and so on.
Thus, the proofs generated during the reuse process are
only those generated by the corresponding B mechanisms.

For each mechanism, we present an implementation in
B and its limits. We refer to the 
example presented in Sect. 2.2. The B
specification is introduced step by step, the complete specification is 
given in \cite{frederic}.

The last subsection presents a summary of the proof activity
generated by the approach.

\subsection{Definition of Patterns}

The way a pattern is defined is strongly dependent on how the reuse
mechanisms are defined.
Ideally, a pattern should be specified by several machines (for
instance, one machine for each class of the UML diagram), all
included in a machine which stands for an interface of the pattern
(see UML-B approach in Sect. 3.1 and \cite{ngu98}).

For technical reasons, some B mechanisms such as refinement
require the use of a single abstract machine. Consequently, a
pattern is specified by a single abstract machine. This is a first
limit of our approach, and perhaps the more annoying one.

In order to implement the composition mechanism (see Sect. 4.3), the 
given sets of
the pattern must be specified as parameters of the machine.
For instance, the {\bfseries Composite} pattern is specified with
the following \texttt{Composite\_Pattern} machine:
\begin{description}
\item[MACHINE] Composite\_Pattern(COMPONENT)
\item[VARIABLES] Component, Composite, Leaf, Father
\item[INVARIANT]
\begin{description}
\item
\item Component $\subseteq$ COMPONENT $\wedge$
\item Composite $\subseteq$ Component $\wedge$
\item Leaf $\subseteq$ Component $\wedge$
\item Father $\in$ Component $\psurj$ Composite $\wedge$
\item Leaf $\cup$ Composite $=$ Component $\wedge$
\item Leaf $\cap$ Composite $=$ $\emptyset$
\end{description}
\item ...
\item[OPERATIONS]
\item children $\longleftarrow$ GetChild(father) $=$ ...
\item cpt $\longleftarrow$ New\_Composite(comp) $=$ ...
\item Add\_Composite(cpt,comp) $=$ ...
\item leaf $\longleftarrow$ New\_Leaf $=$ ...
\item Add\_Leaf(leaf) $=$ ...
\item Remove\_Composite(cpt) $=$ ...
\item Remove\_Leaf(leaf) $=$
\item {\bfseries pre} \\
leaf $\in$ Leaf $\wedge$ \\
leaf $\in$ dom(Father)
\item {\bfseries then} \\
...
\item {\bfseries end}
\item Operation(cpt) $=$ ...
\end{description}

The variables \texttt{Component}, \texttt{Composite} and \texttt{Leaf}
represent three classes, while the variable \texttt{Father}
stands for the association link between the UML classes \texttt{Component}
and \texttt{Composite} (see Sect. 2.1). The types of the
classes and associations are formally specified in the B
invariants.
In the same way, the {\bfseries Resource Allocation} pattern is specified in B.
\begin{description}
\item[MACHINE] Resource\_Allocation(JOBS, CATEGORY, \\
FACILITY, RESOURCE)
\item[SETS] DATE
\item[VARIABLES]
\begin{description}
\item
\item JobOccurrence, When, JobCategory, Resource, ResourceFacility,
\item Job, Requirements, Provides, Allocated
\end{description}
\item[INVARIANT]
\begin{description}
\item
\item JobOccurrence $\subseteq$ JOBS $\wedge$
\item When $\in$ JobOccurence $\rightarrow$ DATE $\wedge$
\item ...
\end{description}
\item[OPERATIONS]
\item Add\_Resource(res) $=$ ...
\item Remove\_Resource(res) $=$
\item {\bfseries pre} \\
res $\in$ RESOURCE
\item {\bfseries then} \\
Resource $:=$ Resource $-$ $\{$res$\}$ $\|$ \\
Provides $:=$ $\{$res$\}$ $\ndres$ Provides $\|$ \\
Allocated $:=$ Allocated $\nrres$ $\{$res$\}$
\item {\bfseries end}
\item ...
\end{description}

\subsection{Instantiation Mechanism}

The instantiation mechanism is implemented in B by the inclusion
of machines. The machine corresponding to the pattern is included
in the machine corresponding to the instantiation of the pattern.

In the example, the machine \texttt{Directory\_Renaming} includes the 
machine of
the {\bfseries Composite} pattern. The given sets, defined as parameters 
of the machine, can be
renamed by instantiation of the parameters. In our example, 
\texttt{COMPONENT} is renamed by \texttt{ELEMENT}:
\begin{description}
\item[MACHINE] Directory\_Renaming
\item[SETS] ELEMENT
\item[INCLUDES] Composite\_Pattern(ELEMENT)
\item ...
\end{description}

The {\bfseries DEFINITIONS} clause allows us to rename
variables from the included machine. \texttt{Composite} and \texttt{Leaf} 
are respectively renamed by \texttt{Directory} and \texttt{File}.
\begin{description}
\item[DEFINITIONS]
\begin{description}
\item
\item Directory $==$ Composite
\item File $==$ Leaf
\end{description}
\end{description}

Renaming operations is not so straightforward. There are two cases. Firstly,
if an operation is directly reused without renaming we use the
{\bfseries PROMOTES} clause. Secondly, a renamed operation is specified 
in the {\bfseries
OPERATIONS} clause. It consists of a call statement to the
corresponding pattern operation. In our example, the \texttt{Remove\_Leaf}
operation is renamed by \texttt{Remove\_File}:
\begin{description}
\item[OPERATIONS]
\item Remove\_File(file) $=$
\item {\bfseries pre} \\
file $\in$ File $\wedge$ \\
file $\in$ dom(Father)
\item {\bfseries then} \\
Remove\_Leaf(file) /$\ast$ operation defined in Composition\_Pattern $\ast$/
\item {\bfseries end}
\end{description}
The preconditions must be the same as those of the called
operation, except that the variables and sets are the renamed
ones.

\subsection{Composition Mechanism}

A first step is to precisely define the composition mechanism. We
distinguish three levels of composition according to whether or
not there exist links between the composed patterns. In the three
cases, composition is achieved by the inclusion mechanism of
B: all the machines representing the composed patterns are
included in the machine representing the composition, called the 
composition machine.

\paragraph{Juxtaposition.}

Patterns are composed without defining any link between them. This 
composition is only a juxtaposition of each pattern. We use 
the {\bfseries EXTENDS}
mechanism which allows all the operations of the two composed
patterns to be considered as genuine operations of
the composition machine:
\begin{description}
\item[MACHINE] Composition\_By\_Juxtaposition
\item[SETS] COMPONENT; JOBS; CATEGORY; FACILITY; RESOURCE
\item[EXTENDS]
\begin{description}
\item
\item Composite\_Pattern(COMPONENT),
\item Resource\_Allocation(JOBS, CATEGORY, FACILITY, RESOURCE)
\end{description}
\item ...
\end{description}

\paragraph{Composition with Inter-Patterns Links.}

New relations between variables of the composed patterns can be
added. For instance, a bijection called \texttt{CompRes} may be defined between
\texttt{Component} (from \texttt{Composite\_Pattern}) and \texttt{Resource}
(from \texttt{Resource\_Allocation}).
\begin{description}
\item[MACHINE] Composition\_By\_InterPatterns\_Links
\item[SETS] COMPONENT; JOBS; CATEGORY; FACILITY; RESOURCE
\item[INCLUDES]
\begin{description}
\item
\item Composite\_Pattern(COMPONENT),
\item Resource\_Allocation(JOBS, CATEGORY, FACILITY, RESOURCE)
\end{description}
\item[VARIABLES] CompRes
\item[INVARIANT] CompRes $\in$ Component $\bij$ Resource
\end{description}

According to the type of the new variables, the operations of the
composed patterns involving the linked variables must be modified.
For instance, the operation \texttt{Remove\_Leaf} from 
\texttt{Composite\_Pattern} removes a \texttt{leaf}
from \texttt{Component}. This
operation cannot be only renamed in the machine 
\texttt{Composition\_By\_Inter\-Patterns\_Links}. It must be composed with
the operation \texttt{Remove\_Resource} from \texttt{Resource\_Allocation} 
and with a substitution on the variable \texttt{CompRes} in order to 
preserve the new invariant.
The resulting operation is \texttt{Remove\_Thing\_1}:
\begin{description}
\item Remove\_Thing\_1(thing) $=$
\item {\bfseries pre} \\
thing $\in$ Leaf $\wedge$ \\
thing $\in$ dom(Father)
\item {\bfseries then} \\
Remove\_Leaf(thing) $\|$ \\
Remove\_Resource(CompRes(thing)) $\|$ \\
CompRes $:=$ $\{$thing$\}$ $\ndres$ CompRes
\item {\bfseries end}
\end{description}

A composition mechanism between operations must be defined. It
could be automated provided that each composed pattern
specifies the elementary operations to be composed on each
variable.

Other operations of the composed patterns can be promoted to become 
genuine operations of
the composition machine.

\paragraph{Unification.}

This composition allows some variables of the composed patterns to
be merged. Up to now, only variables corresponding to classes or
associations in the patterns may be unified.
This property is specified in the {\bfseries INVARIANT} clause of
the composition machine. Two variables may be unified only if they
have the same type, that is why we need to define the given sets
as parameters of a pattern machine (see Sect. 4.1).
For example, we can compose the two patterns \texttt{Composite\_Pattern}
and \texttt{Resource\_Allocation} by unifying the variables 
\texttt{Component} and \texttt{Resource}. This yields the following machine:
\begin{description}
\item[MACHINE] Composition\_By\_Unification
\item[SETS] ELEMENT; JOBS; CATEGORY; FACILITY
\item[INCLUDES]
\begin{description}
\item
\item Composite\_Pattern(ELEMENT),
\item Resource\_Allocation(JOBS, CATEGORY, FACILITY, ELEMENT)
\end{description}
\item[INVARIANT] Component $=$ Resource
\end{description}
Let us note that the parameters \texttt{COMPONENT} and \texttt{RESOURCE} are
now replaced by the same set called \texttt{ELEMENT}.

As in the composition by inter-patterns links, operations of the
composed patterns involving the unified variables must be
redefined. The operation \texttt{Remo\-ve\_Leaf} involves the variable
\texttt{Component} which is now unified with \texttt{Resource}. Then the
operation resulting from the composition is:
\begin{description}
\item Remove\_Thing\_2(thing) $=$
\item {\bfseries pre} \\
thing $\in$ Leaf $\wedge$ \\
thing $\in$ dom(Father)
\item {\bfseries then} \\
Remove\_Leaf(thing) $\|$ \\
Remove\_Resource(thing)
\item {\bfseries end}
\end{description}

From a methodological point of view, an instantiation and a
composition can be achieved in one step, before applying the extension 
mechanism.
For the next subsection, we assume that firstly we have composed the 
two patterns 
\texttt{Composite\_Pattern}
and \texttt{Resource\_Allocation} by unifying the variables 
\texttt{Component} and
\texttt{Resource} and secondly instantiated them by renaming elements 
with the names 
used in the class diagram of
Fig. 3. The resulting B machine is called \texttt{Comp\_By\_Unif\_Inst}:
\begin{description}
\item[MACHINE] Comp\_By\_Unif\_Inst
\item[SETS] ELEMENT; JOBS; CATEGORY; FACILITY
\item[INCLUDES]
\begin{description}
\item
\item Composite\_Pattern(ELEMENT),
\item Resource\_Allocation(JOBS, CATEGORY, FACILITY, ELEMENT)
\end{description}
\item[DEFINITIONS]
\begin{description}
\item
\item Directory $==$ Composite;
\item File $==$ Leaf;
\item SecretaryJobs $==$ JobOccurrence;
\item Secretary $==$ JobCategory;
\item FileFacility $==$ ResourceFacility;
\item FileRequirements $==$ Requirements;
\item Element $==$ Resource
\end{description}
\item[INVARIANT] Component $=$ Resource
\item ...
\item[PROMOTES] GetChild
\item[OPERATIONS]
\item dir $\longleftarrow$ Add\_Directory(files) $=$ ...
\item file $\longleftarrow$ Add\_File $=$ ...
\item Remove\_Directory(dir) $=$ ...
\item Remove\_File(file) $=$ ...
\item ...
\end{description}
In this machine, the above-mentioned operation 
\texttt{Remove\_Thing\_2} is renamed by \texttt{Remove\_File}.

\subsection{Extension}

Most often an extension consists in defining new variables,
modifying existing operations to take into account these new
variables and adding new operations, in the result of a
composition and/or an instantiation, specified in a B machine
called the before-extension machine. Two solutions exist to
implement this mechanism in B, with arguments on both sides. The
first solution consists in including the before-extension machine
into a new machine. The main drawback is that we cannot modify an
included operation. Then, if we want to modify an operation of the
before-extension machine, we need to rename it. We prefer using
the refinement mechanism with the idea of adding more details to a
specification.  We will see later the limits of the refinement
mechanism.

Let us take our example: the machine \texttt{Comp\_By\_Unif\_Inst}
which is the before-extension machine is refined by the machine 
\texttt{Extension}. 

\begin{description}
\item[REFINEMENT] Extension
\item[REFINES]  Comp\_By\_Unif\_Inst
\item[SETS] ELEMENT; JOBS; SECRETARY; FILEFACILITY; \\
STATE $=$ $\{$write,protected$\}$
\item[INCLUDES]
\begin{description}
\item
\item Composite\_Pattern(ELEMENT),
\item Resource\_Allocation(JOBS, SECRETARY, FILEFACILITY, \\
ELEMENT)
\end{description}
\item[VARIABLES] State, Usable
\item[INVARIANT]
\begin{description}
\item
\item Usable $\subseteq$ Element $\wedge$
\item State $\in$ Usable $\pfun$ STATE
\end{description}
\item[INITIALISATION] Usable, State $:=$ $\emptyset$,$\emptyset$
\item ...
\end{description}

Existing operations may be extended in only one way: new substitutions using
the new variables can be specified. For instance, \texttt{Remove\_File}
is refined by:
\begin{description}
\item Remove\_File(file) $=$
\item {\bfseries pre} \\
file $\in$ File $\wedge$ \\
file $\in$ dom(Father)
\item {\bfseries then} \\
Remove\_Leaf(file) $\|$ Remove\_Resource(file) $\|$\\
Usable $:=$ Usable $-$ $\{$file$\}$ $\|$
State $:=$ $\{$file$\}$ $\ndres$ State
\item {\bfseries end}
\end{description}

New operations cannot be added during the refinement process. Thus,
in order to define a new operation, two solutions are possible. Either the
before-extension machine is extended with the \textbf{EXTENDS} clause, to add
new operations, specified as ``skip''. Or the before-extension machine is 
modified by adding these new operations also specified as ``skip''. 
For the sake
of concision, we present the second option. Then, in both cases, each new
operation is refined with new substitutions involving only new variables 
that are not related to variables in the gluing invariant and with calls to 
operations of the included machines. The refinement is then correct.

In our example, the new operation \texttt{Protect\_Element} sets the
state of an element to \texttt{protected}. This operation is defined in the
\texttt{Comp\_By\_Unif\_Inst} machine by:
\begin{description}
\item Protect\_Element(el) $=$
\item {\bfseries pre} \\
el $\in$ Element
\item {\bfseries then} \\
skip
\item {\bfseries end}
\end{description}
The operation is then refined in the machine \texttt{Extension} by:
\begin{description}
\item Protect\_Element(el) $=$
\item {\bfseries pre} \\
el $\in$ Element
\item {\bfseries then} \\
State(el) $:=$ protected $\|$
Usable $:=$ Usable $\cup$ $\{$el$\}$
\item {\bfseries end}
\end{description}

\subsection{Summary and Analysis of the Proof Activity}

The machines corresponding to the different patterns 
(\texttt{Composite\_Pattern} and \texttt{Resource\_Allocation}) have been 
proved with the Atelier B \cite{clearsy} (see Tab. 2). We will now present
a summary of the proofs generated by the reuse mechanisms.

The instantiation mechanism does not generate new proof
obligations since nothing new has been specified. By construction,
operations are automatically proved. Thus the proofs of the
machine corresponding to an instantiation are obvious: they have
already been proved in the included machine.

The composition by juxtaposition of different machines gives a
machine which is automatically proved, since nothing new has been specified. 
The compositions with
inter-patterns links and by unification generate proof obligations
which are to be automatically discharged if the composition of
operations has been correctly elaborated.

For the extension mechanism, the new proof obligations concern, on
one hand, the invariants and operations which have been added and,
on the other hand, the refinement of the modified operations.

Table 2 summarises the proofs of 
the example described in Sect. 4.4. {\bfseries nObv} is the
number of obvious proofs generated and trivially discharged by the 
Atelier B. {\bfseries nPO} represents the number of proof obligations (PO) 
to discharge. {\bfseries nAut} represents the number of POs automatically
discharged and {\bfseries nInt} the number of POs interactively discharged.
All the interactive proofs have been discharged.

\begin{table}
\caption{Proofs result}
\begin{center}
\begin{tabular}{l r r r r}
Machines & nObv & nPO & nAut & nInt \\
\hline
\hline
Composite\_Pattern & 32 & 59 & 47 & 12 \\
\hline
Resource\_Allocation & 42 & 16 & 15 & 1 \\
\hline
\hline
Comp\_By\_Unif\_Inst & 43 & 10 & 10 & 0 \\
\hline
Extension\_Machine & 284 & 33 & 23 & 10 \\
\hline
\end{tabular}
\end{center}
\end{table}

For \texttt{Extension\_Machine}, only ten
proofs have been interactively proved. Six proof
obligations are linked to the preservation of the following
invariant:
\begin{description}
\item State $\in$ Usable $\pfun$ STATE
\end{description}
These proof obligations depend on the new specifications
introduced in the refinement: they are linked to the extension
mechanism.

Four proof obligations concern the
refinement of operations. These four proof obligations have been
proved in the same way: their goal is false because one of their
hypotheses is false. Whatever the extension is, this kind of proof
obligations is always generated: the proof obligations do
not depend on the new specifications, but on the refinement
mechanism. Once they have been proved, we can use the
same strategy in order to prove them again in another extension.

In comparison with our approach, we have specified the same 
example (see Fig. 3) directly in B without using patterns. The resulting 
machine, called 
\texttt{Direct\_Example}, has been proved with the Atelier B (see Tab. 3
for the proofs result). The complete specification can be found in 
\cite{frederic}.

\begin{table}
\caption{Proofs result of \texttt{Direct\_Example}}
\begin{center}
\begin{tabular}{l r r r r}
Machines & nObv & nPO & nAut & nInt \\
\hline
\hline
Direct\_Example & 167 & 87 & 67 & 20 \\
\hline
\end{tabular}
\end{center}
\end{table}

For the same example, the \texttt{Direct\_Example} machine obviously 
requires less proof activity than with our pattern reuse method, since
only one abstract specification is used compared to the two specification 
patterns, the before-extension machine and the refinement used in our 
approach. However, if we assume that the two specification patterns are
previously specified and proved, our approach requires only forty-three
POs ($10+33$) which have to be compared to the eighty-seven for the direct 
specification. Moreover, once POs
have been automatically discharged, only ten POs ($0+10$) must be 
interactively
proved in our approach by pattern reuse, compared to the twenty POs 
that must be interactively discharged for the \texttt{Direct\_Example} 
machine.

In conclusion, since four POs are ``reusable'' in our approach, the reuse
of specification patterns to specify the example described in Sect. 2.2
allows us to save fourteen POs to interactively discharge, provided that 
the two specification patterns \texttt{Composite\_Pattern} and 
\texttt{Resource\_Allocation} are previously specified and proved. Let us
note that the \texttt{Direct\_Example} machine is inspired by the result
of the specification by pattern reuse described in Fig. 3. One
would have undoubtedly specified the same example differently and 
consequently the generated proofs could be different.

\section{Conclusion, Limits and Perspectives}

In this paper, we have presented an approach for reusing patterns
with the B method. We have implemented the different mechanisms
linked to the reuse of patterns by using only the B mechanisms. It
is interesting because the mechanisms are formally defined and we can 
benefit from the advantages of the B
method, especially the ``reuse'' of proofs and the tool.
There are two major drawbacks. The first one is that a pattern is
defined in one machine, which can produce a big machine, difficult
to read and maintain. The second one is the obligation of defining
the new operations of an extension before actually applying the
extension mechanism.

Concerning the pattern reuse mechanisms, the composition of
several instances of the same pattern has not been studied. We
also have to precisely define the mechanism of operations
composition. The following necessary step will be the development of 
a tool to assist the designer during the specification of an application 
by pattern reuse. The example presented in this paper is rather simple. 
However, the used patterns are those described in \cite{gamm95}, which are
patterns largely tested in real designs. A more complex example
just involves more patterns but the method presented in the paper is
still applicable, provided that a tool is available.

In this paper, we have introduced the notion of reuse of proofs. The
aim is to define the notion of proof linked to a machine and to
specify the reuse of proofs with the B method. This perspective is
new, since the reuse of proofs is not possible with a formal
method like B. However, such a possibility requires several works
on new examples in order to analyse the consequences on the proof
obligations.

\end{document}